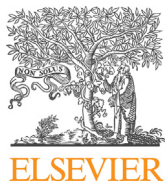
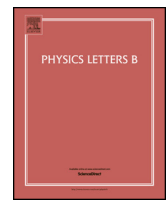

# Non-perturbative rheological behavior of a far-from-equilibrium expanding plasma

Alireza Behtash [a], C.N. Cruz-Camacho [b], Syo Kamata [a], M. Martinez [a,*]

[a] *Department of Physics, North Carolina State University, Raleigh, NC 27695, USA*
[b] *Universidad Nacional de Colombia, Sede Bogotá, Facultad de Ciencias, Departamento de Física, Grupo de Física Nuclear, Carrera 45 N° 26-85, Edificio Uriel Gutiérrez, Bogotá D.C. C.P. 111321, Colombia*

**A B S T R A C T**

For the Bjorken flow we investigate the hydrodynamization of different modes of the one-particle distribution function by analyzing its relativistic kinetic equations. We calculate the constitutive relations of each mode written as a multi-parameter trans-series encoding the non-perturbative dissipative contributions quantified by the Knudsen $Kn$ and inverse Reynolds $Re^{-1}$ numbers. At any given order in the asymptotic expansion of each mode, the transport coefficients get effectively renormalized by summing over all non-perturbative sectors appearing in the trans-series. This gives an effective description of the transport coefficients that provides a new renormalization scheme with an associated renormalization group equation, going beyond the realms of linear response theory. As a result, the renormalized transport coefficients feature a transition to their equilibrium fixed point, which is a neat diagnostics of transient non-Newtonian behavior. As a proof of principle, we verify the predictions of the effective theory with the numerical solutions of their corresponding evolution equations. Our studies strongly suggest that the phenomenological success of fluid dynamics far from local thermal equilibrium is due to the transient rheological behavior of the fluid.

## 1. Introduction

The regime of validity and applicability of relativistic hydrodynamics is linked with the proximity of the system to a local thermal equilibrium. It has, however, been observed in proton-proton collisions [1,2] that phenomenological models based on fluid dynamics give rise to quite satisfactory results, despite the system size being so small [3–5]. Furthermore, various theoretical models [6–18] have shown that hydrodynamics is valid even when large pressure anisotropies are present.

These findings hint at the existence of a new theory for far-from-equilibrium fluid dynamics [19–21]. Although little is known of this theory, it has been identified that the onset of hydrodynamics depends on the decay of new degrees of freedom, the so-called non-hydrodynamic modes [22–26], whose nature is purely non-perturbative [18,19,22,23,25,27–35]. To this end, it is necessary to study how the nonlinear microscopic dynamics between different modes influences its macroscopic response. But a full-fledged study of non-equilibrium physics is not practically possible unless an effective truncation scheme is instead proposed. Thus, the challenge is to identify a tractable truncation scheme which captures the relevant non-equilibrium dynamics while being simple enough to perform concrete calculations. This task requires creating new computational tools and techniques, which is not otherwise achievable via the usual linear response theory.

In this Letter, we develop a new approach to studying hydrodynamization and nonlinear transport processes, which goes beyond linear response regime. Our treatment is highly nonlinear in the Knudsen and inverse Reynolds numbers, and thus, appropriate for the study of far-from-equilibrium dynamics and understanding the non-perturbative information of short-lived non-hydrodynamic modes. With the help of *trans-asymptotic matching* [36], we calculate the sum of short-lived transient non-hydrodynamic modes at a fixed order in the gradient expansion, and the transport coefficients thereby become dynamical quantities evolving with time toward their respective asymptotic equilibrium values obtained by linear response theory. Knowing the late-time (IR) behavior of the fluid, we back-trace the flow of transport coefficients all the way into the early-time (UV) by including as many

* Corresponding author.
*E-mail addresses:* abehtas@ncsu.edu (A. Behtash), cncruzc@gmail.com (C.N. Cruz-Camacho), skamata@ncsu.edu (S. Kamata), mmarti11@ncsu.edu (M. Martinez).

https://doi.org/10.1016/j.physletb.2019.134914
0370-2693/© 2019 The Authors. Published by Elsevier B.V. This is an open access article under the CC BY license (http://creativecommons.org/licenses/by/4.0/). Funded by SCOAP³.

non-perturbative contributions as possible, to the transient non-hydrodynamic modes, which naturally build up the formal exponential solutions to the fluid evolution equations. This consequently provides renormalization group (RG) equations governing the RG flows of the dynamical transport coefficients, which by construction, converge to the correct values as the fluid reaches the thermal equilibrium. This simple situation is analogous to the case of non-Newtonian fluids [37], where transport coefficients such as shear viscosity, become nonlinear functions of the gradient velocity and thus, their evolution is dictated by the macroscopic deformation history of the fluid.[1]

We consider an expanding conformal plasma undergoing Bjorken flow. The microscopic description of this system is governed by the relativistic Boltzmann equation within the relaxation time approximation (RTA-BE). The problem of solving the RTA-BE is transformed into determining the evolution of the moments through their nonlinear kinetic equations. The solutions of the moments can be written in terms of a *multi-parameter trans-series* [36], which correctly yields the information of the attracting section around the thermal equilibrium fixed point for the gas at late time, otherwise known as the "attractor". Being able to go beyond the attracting region distinguishes the predictions of linear response theory and the physics of non-equilibrated fluid dynamics.

## 2. Kinetic model

The $ISO(2) \otimes SO(1,1) \otimes \mathbb{Z}_2$ symmetry group of the Bjorken flow becomes explicitly manifest in the Milne coordinates $x^\mu = (\tau, x, y, \varsigma)$, with longitudinal proper time $\tau = \sqrt{t^2 - z^2}$ and space-time rapidity $\varsigma = \tanh^{-1}(z/t)$, where the metric is $g_{\mu\nu} = \mathrm{diag}(-1, 1, 1, \tau^2)$. The four-momentum of particles is decomposed like $p^\mu = -(u \cdot p) u^\mu + \Xi^{\mu\nu} p_\nu + (l \cdot p) l^\mu$ where the fluid velocity is $u^\mu = (1,0,0,0)$, $l^\mu = (0,0,0,\tau^{-1})$ and $\Xi^{\mu\nu} = g^{\mu\nu} + u^\mu u^\nu - l^\mu l^\nu$ [38]. In the Milne coordinates the Bjorken flow symmetries reduce the RTA-BE to a simple relaxation type equation [39]

$$\partial_\tau f_{\boldsymbol{p}} = -\frac{1}{\tau_r(\tau)} \left[ f_{\boldsymbol{p}} - f_{eq.} \right]. \tag{1}$$

Here $f_{\boldsymbol{p}} = f(\tau, p_T, (p_\varsigma/\tau))$, where the transverse momentum is given by $p_T = |p_\mu \Xi^{\mu\nu} p_\nu|$ and the longitudinal momentum along the $\varsigma$-direction is $p_\varsigma$. We consider $f_{eq.}(p^\tau/T(\tau)) = e^{-p^\tau/T(\tau)}$ where $p^\tau = \sqrt{p_T^2 + (p_\varsigma/\tau)^2}$ is the energy of the particle, and $T(\tau)$ represents the local temperature. For massless particles, the relaxation time scale is $\tau_r(\tau) = \theta_0/T(\tau)$, with $\theta_0 = 5(\eta/s)_0$ [40], and $(\eta/s)_0$ being the equilibrium value of the shear viscosity over entropy ratio.

The nonlinear relaxation process described by the Boltzmann equation can be understood in terms of the evolution of the moments of the distribution function [41]. Thus, we use the following ansatz for $f_{\boldsymbol{k}}$[2]

$$f_{\boldsymbol{p}} = f_{eq.} \sum_{n=0}^{N} \sum_{l=0}^{L} c_{nl}(\tau) \mathcal{P}_{2l}\left(\frac{p_\varsigma}{\tau p^\tau}\right) \mathcal{L}_n^{(3)}\left(\frac{p^\tau}{T}\right), \tag{2}$$

where $\mathcal{L}_n^{(3)}$ and $\mathcal{P}_{2l}$ denote the generalized Laguerre and Legendre polynomials, respectively. From this expression the moments $c_{nl}$ are directly read as

$$c_{nl}(\tau) = \frac{A_{nl}}{T^4(\tau)} \left\langle (p^\tau)^2 \, P_{2l}\left(\frac{p_\varsigma}{\tau p^\tau}\right) \mathcal{L}_n^{(3)}\left(\frac{p^\tau}{T}\right) \right\rangle, \tag{3}$$

where $A_{nl} = 2\pi^2 (4l+1) \frac{n!}{(n+3)!}$ and $\langle \cdots \rangle \equiv \int_{\boldsymbol{p}} \cdots f_{\boldsymbol{p}}$. The hydrodynamic equilibrium is given by $c_{nl}^{eq.} = \delta_{n0}\delta_{l0}$. Here, we only study the low energy modes, i.e. $n = 0$, and thus we denote $c_{0l}$ by $c_l$.[3] Furthermore, we assume that the initial values of $c_{nl}$ are obtained from the RS distribution function ansatz $f_0 = f_{eq.}(\sqrt{p_T^2 + (1+\xi_0)(p_\varsigma/\tau_0)^2}/[R(\xi_0)^{-1/4}T_0])$ [42] where $\xi_0$ measures the momentum anisotropy and $R(\xi_0)$ is given by Eq. (7a) in [43].

For the ansatz (2) the energy-momentum tensor is $T^{\mu\nu} = \langle p^\mu p^\nu \rangle = \epsilon u^\mu u^\nu + \Xi^{\mu\nu} P_T + l^\mu l^\nu P_L$ where the energy density $\epsilon$, the transverse ($P_T$) and longitudinal ($P_L$) pressures are given as follows:

$$\epsilon = \langle (u \cdot p)^2 \rangle = \frac{3 c_0}{\pi^2} T^4, \quad P_T = \langle p_T^2 \rangle = \epsilon \left(\tfrac{1}{3} - \tfrac{1}{15} c_1\right),$$
$$P_L = \langle (p_\varsigma/\tau)^2 \rangle = \epsilon \left(\tfrac{1}{3} + \tfrac{2}{15} c_1\right). \tag{4}$$

The Landau matching condition implies $c_0 \equiv 1$. The only independent component of the normalized effective shear tensor is $\bar{\pi} = \pi_\varsigma^\varsigma / \epsilon = 2/15 \, c_1$.

## 3. The formal trans-series solution

In a scale invariant system governed by (1), an asymptotic series solution can be expanded in terms of the dimensionless parameter $w^{-1} := (\tau T(\tau))^{-1}$ [22]. Then, the underlying non-autonomous dynamical system consisting of evolution equations of the moments and temperature has its dimensions $d = L + 1$ (with $l = 1, \ldots, L$) reduced by one. Furthermore, $w$ can be interpreted as an 'energy' scale for the renormalization scheme arising naturally from summing all the non-perturbative sectors around the (perturbative) asymptotic expansion of the modes $c_l$. Hence, preparing a dynamical system for the modes $c_l(w)$ automatically amounts to having a renormalization group equation on the phase space of the Bjorken flow built out of these modes. These flows emanate from the vicinity of a UV point discussed below and converge to $c_l^{eq}$, alternatively referred to as the IR fixed point due to being reached when $w \to \infty$ (or, equivalently, $Kn, Re^{-1} \to 0$).

The evolution equations of the moments $c_l$ obtained from the conservation laws and the RTA-BE (1) (see Appendix A) can be written as

$$\left(1 - \frac{c_1}{20}\right)\frac{d\tilde{\boldsymbol{c}}}{dw} + \hat{\Lambda}\tilde{\boldsymbol{c}} + \frac{1}{w}\hat{\mathcal{B}}_D \tilde{\boldsymbol{c}} - \frac{c_1}{5w}\tilde{\boldsymbol{c}} + \frac{3}{2w}\tilde{\boldsymbol{\gamma}} = 0, \tag{5}$$

where $\hat{\mathcal{B}}_D = U\hat{\mathcal{B}}U^{-1} = \mathrm{diag}(b_1, \cdots, b_L) \in \mathbb{C}^L$, $\tilde{\boldsymbol{c}} = U\boldsymbol{c}$, and $\tilde{\boldsymbol{\gamma}} = U\boldsymbol{\gamma}$ with $\boldsymbol{\gamma} = (\tfrac{8}{3}, \ldots, 0)^\top$. Note that we can reproduce $c_1$ in (5) by $c_1 = \sum_{l=1}^{L} U_{1l}^{-1} \tilde{c}_l$ and $c_l$ are the components of $\boldsymbol{c}$. This system is $L$-dimensional, rank 1, and level-1 vector differential equation in a neighborhood of an irregular singularity, $w = \infty$. In the analysis of asymptotic solutions of (5), two terms play a crucial role: **(i)** the coefficient of order $\mathcal{O}(w^0 \tilde{\mathbf{c}})$ term, i.e. the diagonal matrix of *Lyapunov exponents* $\hat{\Lambda}$ determining the rates at which exponential stability is reached at equilibrium in $L$ directions of the flow space, and **(ii)** the coefficient of order $\mathcal{O}(w^{-1} \tilde{\mathbf{c}})$, i.e. the (diagonalized) *mode-to-mode coupling* matrix $\hat{\mathcal{B}}_D$ that shows how each mode $c_l$ is coupled to the neighboring modes. Knowing these two elements paves the way to recasting the dynamical system into the form (5) specifically prepared for asymptotic analysis shown first in generic nonlinear rank-1 systems of ODEs in Ref. [44]. Since $\hat{\mathcal{B}}$ is not diagonal as it stands, the *prepared* form of the dynamical system was written in the basis of $\tilde{\mathbf{c}}$ that diagonalizes it. Notice that in the RTA there is only one parameter $\theta_0$ and thus, the Lyapunov exponents are all identically maximal and given by $\hat{\Lambda} = \frac{3}{2\theta_0} \hat{\mathbb{1}}$.

---

[1] A *perturbative* formulation of relativistic fluid dynamics that goes beyond the IS formalism, has been advocated in the past by different authors [52–65], where higher-order dissipative corrections were included. Using such a perturbative resummation, some of the authors of these works [56–61,63,65,66] realized the non-Newtonian nature of the transport coefficients. Nonetheless, none of these approaches studied the non-perturbative contributions of non-hydrodynamic modes at a fixed order in the gradient expansion, which is the main message in this Letter.

[2] Similar ansatz has been used in Lattice Boltzmann methods for relativistic systems [67].

[3] In a recent publication of three of us [68] we have studied what happens with the hydrodynamization of moments $c_{nl}$ when $n, l > 0$ entering in our ansatz (2). One of the major findings in our newest work is that the first dissipative correction in the Knudsen number to the distribution function is not only determined by the well-known effective shear viscous term but also a new high-energy non-hydrodynamic mode $c_{11}$. It is also demonstrated that the survival of this new mode is intrinsically related to the nonlinear mode-to-mode coupling with the shear viscous term.

In building the matrix $U$ it is appropriate to choose the eigenvectors of $\hat{\mathcal{B}}$ in such a way that $U_{Ll}^{-1} = 1$ for any $l$. Doing so yields $b_{2l-1} = b_{2l}^*$ for $l = 1, \ldots, [L/2]$ where $[\bullet]$ is the floor function and $b_L \in \mathbb{R}$ for odd $L$. Also important is to state that rescaling the eigenvectors immediately corresponds to changing the expansion parameters in the trans-series ansatz defined below and for the sake of consistency, we will stick to the above condition for the eigenvectors from now on. Finally, let us write down the trans-series ansatz for $\tilde{c}_l$:

$$\tilde{c}_l(w) = \sum_{\mathbf{n} \geq \mathbf{0}; |\mathbf{n}| \geq 0}^{\infty} \boldsymbol{\sigma}^{\mathbf{n}} \boldsymbol{\zeta}^{\mathbf{n}}(w) \sum_{k=0}^{\infty} \tilde{u}_{l,k}^{(\mathbf{n})} w^{-k}. \tag{6}$$

Here, $\boldsymbol{\zeta}^{\mathbf{n}}(w) = \exp(-(\mathbf{n} \cdot \mathbf{S})w) w^{\mathbf{n} \cdot \tilde{\mathbf{b}}} = \prod_{l=1}^{L} (\zeta_l(w))^{n_l}$ where $\zeta_l(w) = \exp(-S_l w) w^{\tilde{b}_l}$, $\boldsymbol{\sigma}^{\mathbf{n}} \in \mathbb{C}^L$ are the product of expansion parameters $\prod_{l=1}^{L} \sigma_l^{n_l}$, and $n_1, \ldots, n_L$ are non-negative integers. The dot indicates the inner product, $\mathbf{A} \cdot \mathbf{B} = \sum_{l=1}^{L} A_l B_l$. It is better to exploit vector expressions for the non-perturbative sector number aka *trans-monomial* order $\mathbf{n} = (n_1, \ldots, n_L)$, the anomalous dimension of the pseudo-modes $\tilde{c}_l$, $\tilde{\mathbf{b}} = (\tilde{b}_1, \ldots, \tilde{b}_L)$, and the Lyapunov exponents $\mathbf{S} = (S_1, \ldots, S_L)$.

By substituting the ansatz (6) in the differential equation (5), one can obtain a recursive relations for the coefficients as

$$\boldsymbol{\sigma}^{\mathbf{n}} \boldsymbol{\zeta}^{\mathbf{n}} \left[ (\mathbf{n} \cdot \tilde{\mathbf{b}} + b_l - k) \tilde{u}_{l,k}^{(\mathbf{n})} + (\frac{3}{2\theta_0} - \mathbf{n} \cdot \mathbf{S}) \tilde{u}_{l,k+1}^{(\mathbf{n})} \right] + \frac{3}{2} \tilde{\gamma}_l \delta_{k,0} \delta_{\mathbf{n},\mathbf{0}}$$

$$- \frac{\boldsymbol{\sigma}^{\mathbf{n}} \boldsymbol{\zeta}^{\mathbf{n}}}{20} \sum_{l'=1}^{L} \sum_{\mathbf{n}' \geq \mathbf{0}; |\mathbf{n}'| \geq 0}^{\mathbf{n}} U_{1l'}^{-1} \left[ \sum_{k'=0}^{k} (\mathbf{n}' \cdot \tilde{\mathbf{b}} + 4 - k') \tilde{u}_{l',k-k'}^{(\mathbf{n}-\mathbf{n}')} \tilde{u}_{l,k'}^{(\mathbf{n}')} \right.$$

$$\left. - \mathbf{n}' \cdot \mathbf{S} \sum_{k'=0}^{k+1} \tilde{u}_{l',k-k'+1}^{(\mathbf{n}-\mathbf{n}')} \tilde{u}_{l,k'}^{(\mathbf{n}')} \right] = 0, \tag{7}$$

where $\delta_{\mathbf{n},\mathbf{0}} = \delta_{n_1,0} \cdots \delta_{n_L,0}$. With a little bit of algebra, Eq. (23) shows that the modes $c_l(w)$ go like $\mathcal{O}(w^{-l})$, resulting in $\tilde{u}_{l,0}^{(\mathbf{0})} = 0$ for any $l$. So physically, $c_1$ is the slowest mode that contributes to the observables in (4). The zeroth order term in the first non-perturbative sector ($|\mathbf{n}| = 1$) is always a constant normalizing the trans-series expansion parameters $\sigma_l$. For convenience, we choose to set $\tilde{u}_{l,0}^{(\mathbf{n})} = 1$ for $n_{l'} = \delta_{l',l}$. Other trans-series coefficients can be recursively determined order by order in both $k$ and $\mathbf{n}$ in (7).

In addition, at $|\mathbf{n}| = 1$ order Eq. (7) gives $S_l = \frac{3}{2\theta_0}$ and $\tilde{b}_l = -b_l + \frac{1}{5}$. For example, at first two truncation orders $L = 1, 2$ the anomalous dimensions are given as $\tilde{b}_1 = -\frac{18}{35}$, $\tilde{b}_{1,2} = -\frac{1}{110}(53 \pm i\sqrt{10655})$, respectively. It is notable that, physically, all $c_l$ must be real-valued but $\tilde{c}_l$ are in general complex, justifying the name 'pseudo-modes'. The reality condition for $c_l$ is related to the choice of the initial values (expansion parameters) $\sigma_l$, which imposes the constraints $\sigma_{2l-1} = \sigma_{2l}^* \in \mathbb{C}$ for $l = 1, \ldots, [L/2]$, and $\sigma_L \in \mathbb{R}$ if $L$ is odd. Every flow line that evolves to the attractor of the equilibrium fixed point as $w \to \infty$, can be constructed from the set of asymptotic solutions to the dynamical system (5) for some $\sigma_l$. The allowed set of $\sigma_l$ shapes the basin of attraction at any $L$.

## 4. Initial data and UV information

Determining the $l$-th expansion parameter $\sigma_l$ is the most challenging task related to Eq. (6). This partly goes back to the fact that there are $L$ equations involved in the system of ODEs of (5), whereas the exact numerical expression of (3) is always prepared by only three initial values: $\tau_0$, $T_0$ and the parameter $\xi_0$. Among these constants, the search for $\xi_0$-dependence of $\sigma_l$ is more difficult as the other two explicitly enter the trans-series via Écalle time $w$ itself. In order to obtain $\sigma_l$, we use the least squares optimization to fit the trans-series of all the modes involved at order $L$ simultaneously on the Stokes line. One, however, needs to be aware of "Stokes phenomenon", and the UV data that eventually dictate the fate of flow lines.

First, we have to note that capturing the divergence of original asymptotic expansion ($\mathbf{n} = \mathbf{0}$) in (6) is done by initially Borel transforming the series. This yields some sort of singularity along the physically relevant Stokes line, that is, $\Im(\mathbf{n} \cdot \mathbf{S} w) = 0$. Then, the Borel resummation of the original asymptotic series picks up a residue or discontinuity around the Stokes line (given more explicitly by Écalle's bridge equation [45]). Imposing the reality condition, one concludes that the overall constant appearing in the imaginary piece, aka "Stokes constant", has to cancel the imaginary part of the $\sigma_l$. For $L = 1$, it is shown that this constant is $\Im(\sigma_1) = -\pi C_0 \theta_0^{-\tilde{b}_1}$ with the overall fitting factor of $C_0 \approx 0.4898$.

Second, the flow lines heading to the IR should always take values in the basin of attraction, on the boundary of which lies the UV points [18]. These are actually the RG flows that unveil the general behavior of the modes $c_l$ in kinetic theory and consequent renormalization of transport coefficients far from equilibrium. In the context of RTA Boltzmann theory, the RG time should appropriately be chosen as $t = \log(\rho)$ with $\rho = \tau T_0/\theta_0$, and $\tau_0 T_0/\theta_0$ is a UV cutoff. Then, for $L = 1$, we get the following autonomous system of equations describing the RG flows in the 3d macroscopic Bjorken phase space:

$$\frac{dT}{dt} = -\frac{1}{3} T - \frac{1}{30} T c_1, \tag{8a}$$

$$\frac{dc_1}{dt} = -\frac{8}{3} - \frac{10}{21} c_1 + \frac{2}{15} c_1^2 - (T/T_0) c_1 \rho, \tag{8b}$$

$$\frac{d\rho}{dt} = \rho. \tag{8c}$$

Unless $\tau_0 = T_0 = 0$ (while keeping $\theta_0$ finite), the theory does not have a UV completion. At this limit, there are two UV fixed points: $(\rho, T, c_1)_{UV} = (0, 0, -3.03)$, an index-2 (weakly unstable) saddle point, and $(0, 0, 6.60)$ which is an index-1 (strongly unstable) saddle point shown in Fig. 1. The index represents the number of attracting directions that lie on the $T$-$c_1$ plane while the orthogonal direction $\rho$ is obviously repelling. Considering the full theory $L = \infty$, the UV points move into the physical domain and become $(\rho, T, \mathbf{c})_{UV} = \left(0, 0, (-2.5, \{(-1)^l (4l+1) \binom{2l}{l} 4^{-l}\}_{l=2}^{\infty})\right)$ and $\left(0, 0, (5, \{4l+1\}_{l=2}^{\infty})\right)$, of which the former gives the "free-streaming" limit ($\xi_0 \to \infty$). The latter explains a recently found solution in Ref. [46]. At this point, the longitudinal pressure is maximum $P_L = \epsilon$, which in turn causes a huge amount of pressure along the $z$-direction due to $\xi_0 \to -1$, while there is a slowdown in the transverse direction because $P_T = 0$. For a flow emanating from the neighborhood of this point, there is a local relaxation happening at early times, an indication of more attracting directions in the UV, as is obvious from the larger index of the saddle point $A$ (free-streaming) compared to that of the saddle point $B$ in Fig. 1. Thus, we suitably define $\Xi_0 := 1/(1 + \xi_0)$ that measures the proximity of the initial point of flow lines to either UV fixed points. The smaller (or larger) $\Xi_0$, the closer flow lines will be to the saddle points $A$ (or $B$).

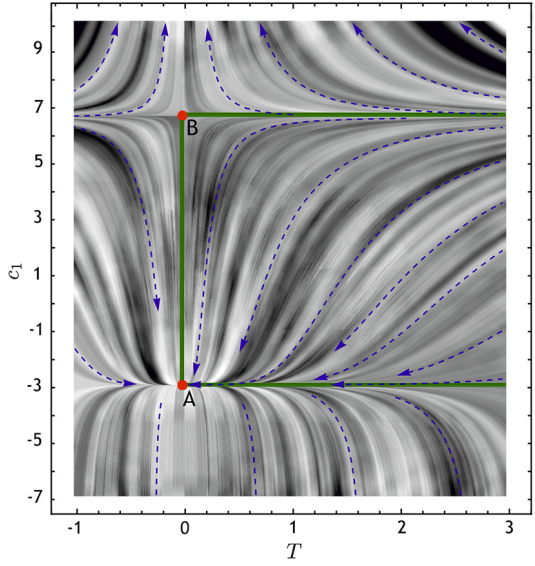

**Fig. 1.** Phase portrait of the Bjorken flow at $\rho \sim 0$ ($w \sim 0$). The saddle points $A$, $B$ show the UV fixed points, which are of index-2 and index-1 type, respectively. $A$ is the free-streaming point. The depth of diagram is the $\rho$-direction. The green lines indicate the boundary of the invariant space of physical domain in the $T$-$c_1$ plane at early times, evolving toward the equilibrium fixed point $(\infty, 0, 0)$ in the 3d phase space for $L = 1$.

## 5. Non-perturbative contributions renormalize the transport coefficients

To find the renormalized coefficients, we straightforwardly do the sum over $\mathbf{n}$ in (7) that leads to the master recursive relation

$$\tilde{\mathbf{b}} \cdot \hat{\boldsymbol{\xi}} \tilde{C}_{l,k} + (b_l - k) \tilde{C}_{l,k} - \mathbf{S} \cdot \hat{\boldsymbol{\xi}} \tilde{C}_{l,k+1} + \frac{3}{2\theta_0} \tilde{C}_{l,k+1} + \frac{3}{2} \tilde{\gamma}_l \delta_{k,0}$$
$$- \frac{1}{20} \sum_{l'=1}^{L} U_{1l'}^{-1} \left[ \sum_{k'=0}^{k} \tilde{C}_{l',k-k'} \left( \tilde{\mathbf{b}} \cdot \hat{\boldsymbol{\xi}} \tilde{C}_{l,k'} + (4 - k') \tilde{C}_{l,k'} \right) \right.$$
$$\left. - \sum_{k'=0}^{k+1} \tilde{C}_{l',k-k'+1} \mathbf{S} \cdot \hat{\boldsymbol{\xi}} \tilde{C}_{l,k'} \right] = 0, \quad (9)$$

where $\tilde{C}_{l,k}(\sigma \zeta(w)) = \sum_{\mathbf{n} \geq 0}^{\infty} \sigma^{\mathbf{n}} \zeta^{\mathbf{n}}(w) \tilde{u}_{l,k}^{(\mathbf{n})}$, $\hat{\zeta}_l = \partial_{\log \zeta_l}$. Notice that (9) is composed of $L$ nonlinear PDEs for every $k$, and it is in general difficult to solve it analytically for $L > 1$ [47]. Nonetheless, it is exactly solvable for $L = 1$ at any order $k$. The exact results for $k = 0, 1$ are given by $\tilde{C}_{1,0}(\sigma_1 \zeta_1) = -20 W_\zeta$ and $\tilde{C}_{1,1}(\sigma_1 \zeta_1) = -\frac{8}{15} \theta_0 (50 W_\zeta^3 + 105 W_\zeta^2 + 36 W_\zeta + 5)/(W_\zeta + 1)$ respectively, where $W_\zeta = W(-\sigma_1 \zeta_1/20)$ is the Lambert-$W$ function. Fig. 2 compares the bare and renormalized $c_1$ for different initial conditions for $L = 1$, and the agreement between the exact numerical and trans-series solutions is overall remarkable.

We may directly extract the renormalized transport coefficients from the near-IR asymptotic expansion of non-hydrodynamic modes $c_l$ found in [48] all the way into near-UV region by including the exponentially small contributions coming from the nonlinearities introduced in Eq. (5) by mode-to-mode coupling terms. Here, we have to pay attention to two important characteristics of an RG flow in connection with this simple picture:

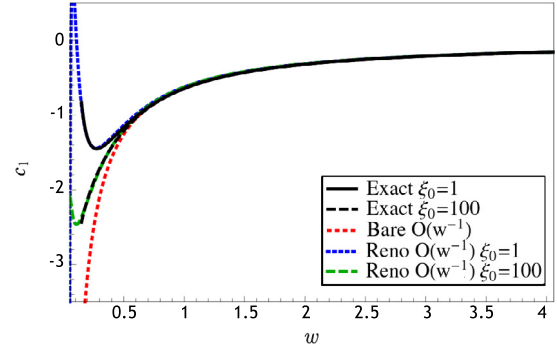

**Fig. 2.** $c_1$ vs. $w$ for $\xi_0 = 1, 100$. The bare and renormalized values of $c_1$ are taken into account up to $O(w^{-1})$. The exact lines are given by numerically solving Eq. (5) for $L = 1$ by setting an initial value of $c_1(w_0)$ with $w_0 = 0.15$ obtained by the numerical analysis of Eq. (3) for $\xi_0 = 1, 100$. For $\xi_0 = 100$, $c_1$ emanates from the neighborhood of the free-streaming point, and thus there is an immediate escape to the hydrodynamic equilibrium without any early-time relaxation. A local minimum (relaxation point) exists that is apparent at $w \sim 0.3$ for $\xi_0 = 1$ in this figure.

**(1)** The cutoff $\tau_0 T_0 / \theta_0$ for a fixed finite $\theta_0$, is merely a physical bound on the near-UV theory such that, as in Fig. 1, the flows are initiated away from the $T$-$c_1$ plane and well inside the basin of attraction of the IR fixed point. **(2)** The anomalous dimensions $\tilde{b}_l$ and Lyapunov exponents $S_l$ are both the IR data appearing in the trans-series of each non-hydrodynamic mode. Along an RG flow initiated at some near-UV point when $\log(w_0/\theta_0) > -\infty$ or $\log \rho > -\infty$ in the basin of attraction, these two parameters never flow with the RG time. In other words, these data are invariant under an RG transformation of energy-momentum tensor governing the dynamical system (5), which pushes forward the flow lines starting near a fixed point and flowing to another fixed point.

It is, therefore, sought to find a non-perturbative approach to determine the far-from-equilibrium transport coefficients from a rheologic interpretation of the fluid. The deformation history of the fluid is traced in the transport coefficients by considering their nonlinear functional dependence on the velocity gradient tensor, e.g. $\eta \equiv \eta(\sigma^{\mu\nu})$, which is often studied at a perturbative level in non-Newtonian fluid dynamics. This has given rise to some interesting physical phenomena known as shear thinning and thickening [37].

The second order gradient expansion of the distribution function [48–51] determines the asymptotic form of $c_1$ to be

$$c_1 = -\frac{40}{3} w^{-1} (\eta/s)_0 - \frac{80}{9} T w^{-2} \left( \tau_{\pi,0} (\eta/s)_0 - (\lambda_1/s)_0 \right) + \ldots \quad (10)$$

Thus, by comparing this result with the one obtained from the trans-series of $c_1$ (for $L = 1$), that is $c_1 = \sum_{k=0}^{\infty} F_{1,k}(w) w^{-k}$, we see that the first non-perturbatively corrected asymptotic term of order $w^{-1}$ gives $\eta/s = -\frac{3}{40} F_{1,1}(w)$, where $\mathbf{F}_k = U^{-1} \tilde{\mathbf{C}}_k$ from (9). In a similar fashion, one can obtain $T(\lambda_1/s - \tau_\pi \eta/s) = \frac{9}{80} F_{1,2}(w)$ from $c_1$. The RG flow equation of these transport coefficients can then be derived from the RG flow of moments. Considering a generic observable $\mathcal{O}(w)$ constituting the resummed coefficients $F_1 = \{F_{1,k} | k \in \mathbb{Z}_{\geq 0}\}$, we can write down its ($L = 1, \theta_0 < \infty$) RG flow equation explicitly as

$$\frac{d\mathcal{O}(F_1(w))}{d \log w} = \sum_{k=0}^{\infty} \left( \tilde{b}_1 - \frac{3w}{2\theta_0} \right) \cdot \left[ \hat{\zeta}_1 F_{1,k}(\zeta_1) \right]_{\zeta_1 = \sigma_1 \zeta_1(w)} \Delta_k(w) \quad (11)$$

where $\zeta_1(w) = e^{-\frac{3w}{2\theta_0}} w^{\tilde{b}_1}$, $F_{1,k}(w) \equiv F_{1,k}(\sigma_1 \zeta_1(w))$ and $\Delta_k(w) = \partial \mathcal{O}(F_1(w))/\partial F_{1,k}(w)$. As can be seen, the transport coefficients get a fully non-perturbative expression through $\zeta_1$. By treating $\zeta_1$ as an effectively independent variable from $w$ [36], $F_{1,k}(\zeta_1)$ can be regarded as the coefficients of a power series in $w^{-1}$, i.e. $c_1(\zeta_1, w) = \sum_{k=0}^{\infty} F_{1,k}(\zeta_1) w^{-k}$. Hence, the piece $\hat{\zeta}_1 F_{1,k}$ in (11) is calculated by using the Cauchy integral formula to pick out the $k$th coefficient of $\hat{\zeta}_1 c_1$:

$$\hat{\zeta}_1 F_{1,k}(\zeta_1) = \frac{1}{2\pi i} \oint_{|w| \ll 1} dw\, w^{k-1} \left(\tilde{b}_1 - \frac{3w}{2\theta_0}\right)^{-1} \times \left[\sum_{k'=0}^{\infty} k' F_{1,k'}(\zeta_1) w^{-k'} + \beta_1(c_1(\zeta_1, w), w)\right], \quad (12)$$

where the *beta* function $\beta_1$ is defined through (5) as $\frac{dc_1}{d\log w} = \beta_1(c_1, w)$. Setting $\mathcal{O}(w) = -\frac{3}{40} F_{1,1}$ and $\mathcal{O}(w) = \frac{9}{80} F_{1,2}$ renders the RG flow equation in terms of $\eta/s$ and $T(\lambda_1/s - \tau_\pi \eta/s)$, respectively. For different initial values of $\sigma_1$, Fig. 3 shows the $w$-dependence of $\eta/s$ and $T(\lambda_1/s - \tau_\pi(\eta/s))$. As $w \to \infty$, both observables asymptotically approach to their equilibrium values at the IR fixed point. There is, however, a shear thinning effect in the UV regime due to the early-time free streaming expansion, as seen in Fig. 3, until the interactions kick in and the flow initiates its relaxation onward exponentially in the shear thickening phase. We also observe that in accordance with the second law of thermodynamics, the renormalized $\eta/s$ remains always positive, and its shape is identical (likewise for $T(\lambda_1/s - \tau_\pi(\eta/s))$) regardless of the initial values. There is an overall tendency to merge to the IR equilibrium faster as $\Xi_0 \to 0$, which is again in line with the argument given at the end of last section.

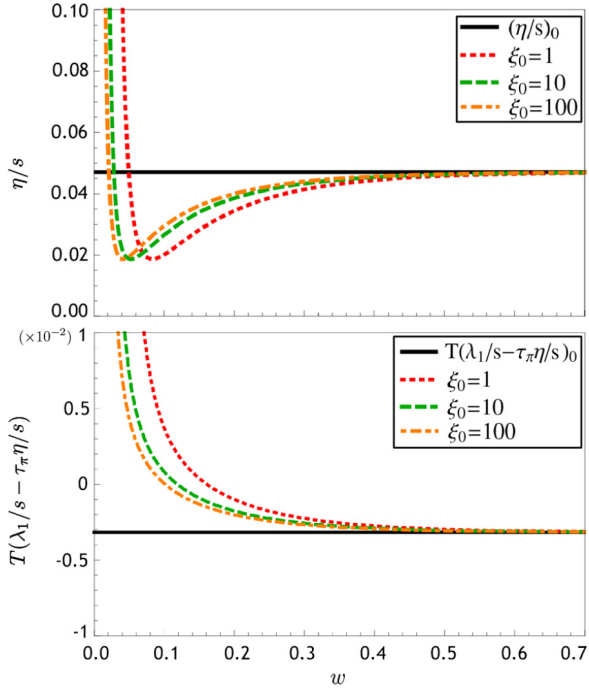

**Fig. 3.** Renormalized $\eta/s$ and $T(\lambda_1/s - \tau_\pi(\eta/s))$ (top and bottom panels respectively) as a function of $w$ for different initial conditions $\xi_0 = \{1, 10, 100\}$ corresponding to the expansion parameters $\sigma_1 = \{1.466, 0.959, 0.791\}$. The values of $\sigma_1$ are chosen after numerical fitting of $c_1$ in Fig. 2.

## 6. Conclusions

We outlined a new method of decomposing the distribution function in terms of non-hydrodynamic modes in a suitable basis, which cast the Boltzmann equation into a dynamical system for a non-equilibrium scale-invariant expanding plasma. We found the UV and IR fixed points of this system and examined its flow lines. It was shown that the exponentially small corrections tracing the nonlinear deformation history of the these flows eventually give a prescription for the renormalization of transport coefficients via trans-asymptotic matching [36]. The short-lived non-hydrodynamic modes give rise to non-Newtonian characteristics of the renormalized transport coefficients, and as a result, hydrodynamics becomes valid in far-from-equilibrium regimes where $Kn$ and $Re^{-1}$ are large. Therefore, we conclude that the success of hydrodynamics in high energy nuclear collisions is intrinsically related with the transient rheological behavior of the fluid.

## Acknowledgements

O. Costin has our sincere gratitude for his patience, extensive discussions and kind explanations of some concepts from his seminal paper. We thank G. Dunne, A. Kemper, M. Spalinski, M. Heller, E. Calzetta, U. Heinz, J. Casalderrey-Solana, A. Muronga, C. Pinilla, T. Schäfer and M. Ünsal for useful discussions. A. B., S. K. and M. M. are supported in part by the US Department of Energy Grant No. DE-FG02-03ER41260, and M. M. is supported by the BEST (Beam Energy Scan Theory) DOE Topical Collaboration.

## Appendix A. Evolution equation of the moments $c_{nl}$

In this supplemental material we briefly expose the derivation of the evolution equation of the moments $c_{nl}$ in (3). $c_{nl}$ can be conveniently written as

$$c_{nl}(\tau) = s_{n,l} \int d^3 p\, A_{n,2l}(\tau)\, f(\tau, p_T, p_\varsigma), \quad (13)$$

where $s_{n,l}$ and $A_{n,2l}$ are respectively given by

$$s_{n,l} = \frac{1}{4\pi}\Gamma(n+1)(4l+1)/\Gamma(n+4), \quad (14a)$$

$$A_{n,2l}(\tau) = \frac{1}{T^4}\frac{E_p}{\tau}\mathcal{L}_n^{(3)}\left(\frac{E_p}{T}\right)\mathcal{P}_{2l}\left(\frac{p_\varsigma}{\tau E_p}\right). \quad (14b)$$

Next, by taking the time derivative of Eq. (13) one obtains

$$\dot{c}_{nl} = \int d^3 p\, s_{n,2l} A_{n,2l}\dot{f} + \int d^3 p\, s_{n,2l}\dot{A}_{n,2l} f, \quad (15)$$

where $\tau$ derivative is denoted by an overdot. The first term on the r.h.s. of (15) can be rewritten as

$$\int d^3 p\, s_{n,2l} A_{n,2l}\dot{f} = -\int d^3 p\, s_{n,2l} A_{n,2l}(f - f^{eq})/\tau_r$$
$$= -\tau_r^{-1}(c_{nl} - \delta_{n0}\delta_{l0}), \quad (16)$$

where use was made of $c_{nl}^{eq} = \delta_{n0}\delta_{l0}$ together with the Boltzmann equation (1). After some algebra, the second term in Eq. (15) reads

$$\int d^3 p\, s_{n,2l}\dot{A}_{n,2l} f = -\tau^{-1}\Big[\alpha_{nl}c_{n,l+1} + \beta_{nl}c_{nl} + \gamma_{nl}c_{n,l-1}$$
$$- n\left(\rho_l c_{n-1,l+1} + \psi_l c_{n-1,l} + \phi_l c_{n-1,l-1}\right)\Big], \quad (17)$$

where the coefficients $\alpha_{nl}, \beta_{nl}, \gamma_{nl}, \rho_l, \psi_l,$ and $\phi_l$ are

$$\alpha_{nl} = \frac{(2+2l)(1+2l)(n+1-2l)}{(4l+3)(4l+5)}, \quad (18a)$$

$$\beta_{nl} = \frac{2l(2l+1)(5+2n)}{3(4l+3)(4l-1)} - \frac{(4+n)c_{01}}{30}, \quad (18b)$$

$$\gamma_{nl} = \frac{(2l+2+n)(2l)(2l-1)}{(4l-3)(4l-1)}, \quad (18c)$$

$$\rho_l = \frac{(2l+1)(2l+2)}{(4l+3)(4l+5)}, \tag{18d}$$

$$\psi_l = \frac{4l(2l+1)}{3(4l+3)(4l-1)} - \frac{c_{01}}{30}, \tag{18e}$$

$$\phi_l = \frac{(2l)(2l-1)}{(4l-3)(4l-1)}. \tag{18f}$$

In Eq. (17) we employed the following properties of Laguerre and Legendre polynomials:

$$(n+1)\mathcal{P}_{n+1}(x) = (2n+1)x\mathcal{P}_n(x) - n\mathcal{P}_{n-1}(x), \tag{19a}$$

$$\frac{(x^2-1)}{n}\frac{d}{dx}\mathcal{P}_n(x) = x\mathcal{P}_n(x) - \mathcal{P}_{n-1}(x), \tag{19b}$$

$$n\mathcal{L}_n^{(\alpha)}(x) = (n+\alpha)\mathcal{L}_{n-1}^{(\alpha)}(x) - x\mathcal{L}_{n-1}^{(\alpha+1)}(x), \tag{19c}$$

$$\frac{d^k}{dx^k}\mathcal{L}_n^{(\alpha)}(x) = \begin{cases} (-1)^k \mathcal{L}_{n-k}^{(\alpha+k)}(x) & \text{if } k \leq n, \\ 0 & \text{otherwise}. \end{cases} \tag{19d}$$

After equating Eqs. (16)-(17) with Eq. (15), one obtains the following evolution equation:

$$\dot{c}_{nl} + \tau^{-1}[\alpha_{nl} c_{n,l+1} + \beta_{nl} c_{nl} + \gamma_{nl} c_{n,l-1} - n(\rho_l c_{n-1,l+1}$$
$$+ \psi_l c_{n-1,l} + \phi_l c_{n-1,l-1})] + \tau_r^{-1} c_{nl} = 0, \tag{20}$$

where $n \geq 0$ and $l > 0$. The energy-momentum conservation law gives the evolution equation of the temperature upon using the matching condition as well as the relations in (4):

$$\dot{\epsilon} + \frac{1}{\tau}(\epsilon + P_L) = 0 \Rightarrow \dot{T} + \frac{T}{3\tau} = -\frac{T c_{01}}{30\tau}. \tag{21}$$

Let us consider only the case $n = 0$ and choose to denote $c_{0l}$ by $c_l$ below. Equations (20) and (3) can be rewritten in terms of the variable $w = \tau T(\tau)$ as

$$\frac{d\log T}{d\log \tau} = -\frac{1}{3}\left(\frac{c_1}{10} + 1\right), \tag{22}$$

$$\left(1 - \frac{c_1}{20}\right)\frac{d\mathbf{c}}{dw} = -\left[\hat{\Lambda}\mathbf{c} + \frac{1}{w}\mathcal{B}\mathbf{c} - \frac{c_1}{5w}\mathbf{c} + \frac{3}{2w}\boldsymbol{\gamma}\right], \tag{23}$$

where $\mathbf{c}(w) = (c_1(w), \cdots, c_l(w))^\top$, $\boldsymbol{\gamma} = (8/3, 0, \cdots, 0)^\top$, $\hat{\Lambda} = 3/(2\theta_0)\hat{\mathbb{1}}$, and

$$\mathcal{B} = \frac{3}{2}\begin{pmatrix} \frac{2}{3}\Omega_1 & \alpha_1 & & & \\ \gamma_2 & \frac{2}{3}\Omega_2 & \alpha_2 & & \\ & \ddots & \ddots & \ddots & \\ & & \gamma_{L-1} & \frac{2}{3}\Omega_{L-1} & \alpha_{L-1} \\ & & & \gamma_L & \frac{2}{3}\Omega_L \end{pmatrix} \tag{24}$$

with $\Omega_l = 5l(2l+1)/[(4l+3)(4l-1)]$. After solving (23) for $c_l$, one gets the temperature $T$ from (22). Finally, the transformation $\tilde{\mathbf{c}} = U\mathbf{c}$ leads us to Eq. (5).

## References

[1] V. Khachatryan, et al., CMS, Phys. Rev. Lett. 116 (2016) 172302, arXiv:1510.03068 [nucl-ex].
[2] G. Aad, et al., ATLAS, Phys. Rev. Lett. 116 (2016) 172301, arXiv:1509.04776 [hep-ex].
[3] R.D. Weller, P. Romatschke, Phys. Lett. B 774 (2017) 351, arXiv:1701.07145 [nucl-th].
[4] K. Werner, I. Karpenko, T. Pierog, Phys. Rev. Lett. 106 (2011) 122004, arXiv:1011.0375 [hep-ph].
[5] P. Bozek, Eur. Phys. J. C 71 (2011) 1530, arXiv:1010.0405 [hep-ph].
[6] A. Kurkela, Y. Zhu, Phys. Rev. Lett. 115 (2015) 182301, arXiv:1506.06647 [hep-ph].
[7] A. Kurkela, E. Lu, Phys. Rev. Lett. 113 (2014) 182301, arXiv:1405.6318 [hep-ph].
[8] R. Critelli, R. Rougemont, J. Noronha, arXiv:1709.03131 [hep-th], 2017.
[9] G.S. Denicol, U.W. Heinz, M. Martinez, J. Noronha, M. Strickland, Phys. Rev. Lett. 113 (2014) 202301, arXiv:1408.5646 [hep-ph].
[10] W. Florkowski, R. Ryblewski, M. Strickland, Nucl. Phys. A 916 (2013) 249, arXiv:1304.0665 [nucl-th].
[11] W. Florkowski, R. Ryblewski, M. Strickland, Phys. Rev. C 88 (2013) 024903, arXiv:1305.7234 [nucl-th].
[12] G.S. Denicol, U.W. Heinz, M. Martinez, J. Noronha, M. Strickland, Phys. Rev. D 90 (2014) 125026, arXiv:1408.7048 [hep-ph].
[13] P.M. Chesler, L.G. Yaffe, Phys. Rev. D 82 (2010) 026006, arXiv:0906.4426 [hep-th].
[14] M.P. Heller, R.A. Janik, P. Witaszczyk, Phys. Rev. Lett. 108 (2012) 201602, arXiv:1103.3452 [hep-th].
[15] B. Wu, P. Romatschke, Int. J. Mod. Phys. C 22 (2011) 1317, arXiv:1108.3715 [hep-th].
[16] W. van der Schee, Phys. Rev. D 87 (2013) 061901, arXiv:1211.2218 [hep-th].
[17] J. Casalderrey-Solana, M.P. Heller, D. Mateos, W. van der Schee, Phys. Rev. Lett. 111 (2013) 181601, arXiv:1305.4919 [hep-th].
[18] A. Behtash, C.N. Cruz-Camacho, M. Martinez, Phys. Rev. D 97 (2018) 044041, arXiv:1711.01745 [hep-th].
[19] P. Romatschke, Phys. Rev. Lett. 120 (2018) 012301, arXiv:1704.08699 [hep-th].
[20] P. Romatschke, U. Romatschke, arXiv:1712.05815 [nucl-th], 2017.
[21] W. Florkowski, M.P. Heller, M. Spalinski, Rep. Prog. Phys. 81 (2018) 046001, arXiv:1707.02282 [hep-ph].
[22] M.P. Heller, M. Spalinski, Phys. Rev. Lett. 115 (2015) 072501, arXiv:1503.07514 [hep-th].
[23] M.P. Heller, V. Svensson, arXiv:1802.08225 [nucl-th], 2018.
[24] P. Romatschke, Eur. Phys. J. C 77 (2017) 21, arXiv:1609.02820 [nucl-th].
[25] M.P. Heller, A. Kurkela, M. Spalinski, Phys. Rev. D 97 (2018) 091503, arXiv:1609.04803 [nucl-th].
[26] M. Spaliński, Phys. Rev. D 94 (2016) 085002, arXiv:1607.06381 [hep-th].
[27] J. Casalderrey-Solana, N.I. Gushterov, B. Meiring, arXiv:1712.02772 [hep-th], 2017.
[28] G.S. Denicol, J. Noronha, Phys. Rev. D 97 (2018) 056021, arXiv:1711.01657 [nucl-th].
[29] P. Romatschke, arXiv:1710.03234 [hep-th], 2017.
[30] M. Strickland, J. Noronha, G. Denicol, Phys. Rev. D 97 (2018) 036020, arXiv:1709.06644 [nucl-th].
[31] G. Basar, G.V. Dunne, Phys. Rev. D 92 (2015) 125011, arXiv:1509.05046 [hep-th].
[32] I. Aniceto, M. Spaliński, Phys. Rev. D 93 (2016) 085008, arXiv:1511.06358 [hep-th].
[33] M. Spaliński, Phys. Lett. B 776 (2018) 468, arXiv:1708.01921 [hep-th].
[34] M. Lublinsky, E. Shuryak, Phys. Rev. C 76 (2007) 021901, arXiv:0704.1647 [hep-ph].
[35] G.S. Denicol, J. Noronha, arXiv:1804.04771 [nucl-th], 2018.
[36] O. Costin, R. Costin, Invent. Math. 145 (2001) 425.
[37] R. Larson, R. Larson, The Structure and Rheology of Complex Fluids, Topics in Chemical Engineering, OUP, USA, 1999.
[38] E. Molnar, H. Niemi, D.H. Rischke, Phys. Rev. D 93 (2016) 114025, arXiv:1602.00573 [nucl-th].
[39] G. Baym, Phys. Lett. B 138 (1984) 18.
[40] G.S. Denicol, J. Noronha, H. Niemi, D.H. Rischke, Phys. Rev. D 83 (2011) 074019, arXiv:1102.4780 [hep-th].
[41] H. Grad, Commun. Pure Appl. Math. 2 (1949) 331.
[42] P. Romatschke, M. Strickland, Phys. Rev. D 68 (2003) 036004, arXiv:hep-ph/0304092.
[43] M. Martinez, M. Strickland, Nucl. Phys. A 848 (2010) 183, arXiv:1007.0889 [nucl-th].
[44] O. Costin, Duke Math. J. 93 (1998) 289.
[45] D. Dorigoni, arXiv:1411.3585 [hep-th], 2014.
[46] A. Dash, A. Jaiswal, Phys. Rev. D 97 (2018) 104005, arXiv:1711.07130 [gr-qc].
[47] A. Behtash, et al., In preparation.
[48] J.-P. Blaizot, L. Yan, J. High Energy Phys. 11 (2017) 161, arXiv:1703.10694 [nucl-th].
[49] J.-P. Blaizot, L. Yan, Phys. Lett. B 780 (2018) 283, arXiv:1712.03856 [nucl-th].
[50] D. Teaney, L. Yan, Phys. Rev. C 89 (2014) 014901, arXiv:1304.3753 [nucl-th].
[51] M.A. York, G.D. Moore, Phys. Rev. D 79 (2009) 054011, arXiv:0811.0729 [hep-ph].
[52] R. Geroch, L. Lindblom, Phys. Rev. D 41 (1990) 1855.
[53] I.-S. Lium, I. Müller, T. Ruggeri, Ann. Phys. 169 (1986) 191.
[54] R. Geroch, L. Lindblom, Ann. Phys. 207 (1991) 394.
[55] A. Muronga, Phys. Rev. C 69 (2004) 034903, arXiv:nucl-th/0309055 [nucl-th].
[56] U.W. Heinz, H. Song, A.K. Chaudhuri, Phys. Rev. C 73 (2006) 034904, arXiv:nucl-th/0510014 [nucl-th].
[57] T. Koide, G.S. Denicol, P. Mota, T. Kodama, Phys. Rev. C 75 (2007) 034909, arXiv:hep-ph/0609117 [hep-ph].
[58] T. Koide, Phys. Rev. E 75 (2007) 060103.
[59] T. Koide, T. Kodama, Phys. Rev. E 78 (2008) 051107.
[60] U.W. Heinz, arXiv:0901.4355 [nucl-th], 2009.

[61] J. Peralta-Ramos, E. Calzetta, Phys. Rev. D 80 (2009) 126002, arXiv:0908.2646.
[62] A. El, Z. Xu, C. Greiner, Phys. Rev. C 81 (2010) 041901, arXiv:0907.4500 [hep-ph].
[63] K. Tsumura, T. Kunihiro, Y. Kikuchi, Physica D 336 (2016) 1, arXiv:1311.7059 [physics.flu-dyn].
[64] A. Jaiswal, Phys. Rev. C 88 (2013) 021903, arXiv:1305.3480 [nucl-th].
[65] A. Harutyunyan, A. Sedrakian, D.H. Rischke, arXiv:1804.08267 [nucl-th], 2018.
[66] K. Tsumura, T. Kunihiro, Eur. Phys. J. A 48 (2012) 162.
[67] P. Romatschke, M. Mendoza, S. Succi, Phys. Rev. C 84 (2011) 034903, arXiv:1106.1093 [nucl-th].
[68] A. Behtash, S. Kamata, M. Martinez, H. Shi, Phys. Rev. D 99 (2019) 116012, arXiv:1901.08632 [hep-th].